\documentstyle[epsfig,12pt]{article}
\newcommand{\be}{\begin{eqnarray}}
\newcommand{\ee}{\end{eqnarray}}

\def\lsim{\mathrel{\rlap{\lower4pt\hbox{\hskip1pt$\sim$}}
    \raise1pt\hbox{$<$}}}               
\def\gsim{\mathrel{\rlap{\lower4pt\hbox{\hskip1pt$\sim$}}
    \raise1pt\hbox{$>$}}}               

\begin{document}

\rightline{\Large{Preprint RM3-TH/99-16}}

\vspace{2cm}

\begin{center}

\LARGE{Target-mass corrections and the Bloom-Gilman duality of the nucleon 
structure function\footnote{\bf To appear in Physics Letters B.}}\\

\vspace{1cm}

\large{Silvano Simula\footnote{E-mail address: simula@roma3.infn.it}}\\

\vspace{0.5cm}

\normalsize{Istituto Nazionale di Fisica Nucleare, Sezione Roma III\\ Via della Vasca Navale 84, I-00146 Roma, Italy}

\end{center}

\vspace{1cm}

\begin{abstract}

\noindent The occurrence of the Bloom-Gilman local duality in the low-order moments of the nucleon structure function is investigated for values of the squared four-momentum transfer $Q^2$ between $\sim 0.5$ and $10 ~ (GeV/c)^2$.  At variance with previous analyses truncated Cornwall-Norton moments, limited to the nucleon-resonance production regions, are considered. The role played by target-mass corrections is illustrated, showing that target-mass effects are necessary (but not sufficient) for producing the observed Bloom-Gilman duality of the nucleon structure function. The possibility of a local duality between the {\em unphysical} region at large values of the Nachtmann variable and the nucleon elastic peak contribution is analyzed. It is found that the proton magnetic form factor extracted assuming local duality is significantly below the experimental data at low and intermediate values of $Q^2$.

\end{abstract}

\vspace{1cm}

PACS numbers: 13.60.Hb, 13.60.Fz, 12.38.-t,12.40.Nn.

\vspace{0.25cm}

Keywords: \parbox[t]{12cm}{nucleon structure function; target-mass corrections; parton-hadron duality; nucleon elastic form factors}

\newpage

\pagestyle{plain}

\section{Introduction}

\indent The investigation of inclusive inelastic lepton scattering off the nucleon (and nuclei) can shed light on the notion of parton-hadron duality. The latter deals with the relation among the physics of the nucleon-resonance production occurring at moderate values of the squared four-momentum transfer, $Q^2$, and the physics of Deep Inelastic Scattering ($DIS$) emerging at high $Q^2$. As it is well known, a local parton-hadron duality was observed empirically by Bloom and Gilman \cite{BG70} in the proton structure function measured at $SLAC$. More precisely, they found that the smooth scaling curve $F_2^p(x')$ characterizing the $DIS$ region at high $Q^2$ represents a good average over the resonance bumps of the inclusive $\nu W_2^p(x', Q^2)$ data  seen at low $Q^2$ for the same values of an improved empirical variable $x' \equiv x / (1 + M^2 x^2 / Q^2)$, where $x$ is the Bjorken scaling variable and $M$ the nucleon mass.

\indent Within $QCD$, a justification of the Bloom-Gilman ($BG$) duality was offered by De Rujula, Georgi and Politzer \cite{RGP77} in terms of the  Cornwall-Norton moments $M_n(Q^2)$ of the nucleon structure function, defined as
 \be
    M_n(Q^2) \equiv \int_0^1 d\xi ~ \xi^n {F_2(\xi, Q^2) \over \xi^2}
    \label{eq:Mn_CN}
 \ee
where $\xi$ is the Nachtmann variable (cf. \cite{GP76})
 \be
    \xi = {2x \over 1 + \sqrt{1 + {4 M^2 x^2 \over Q^2}}}
    \label{eq:csi}
 \ee 
which properly generalizes the empirical $BG$ variable $x'$ [$\xi = x' + O(M^4 x^4 / Q^4)$]. Using the Operator Product Expansion ($OPE$) the authors of Ref. \cite{RGP77} argued that
 \be
    M_n(Q^2) = A_n(Q^2) + \sum_{k=1}^{\infty} \left ( n {\gamma^2
    \over Q^2} \right )^k B_{nk}(Q^2)
    \label{eq:Mn_DGP}
 \ee
where the first term $A_n(Q^2)$  is the result of perturbative $QCD$, while the remaining terms $B_{nk}(Q^2)$ are higher twists related to parton-parton correlations. In Eq. (\ref{eq:Mn_DGP}) the constant $\gamma^2$ represents a scale constant for higher-twist effects. It is clear that a small value of $\gamma^2$ (as argued in Ref. \cite{RGP77} and, more recently, in Ref. \cite{JU95}) makes it possible that for low values of $n$ and for $Q^2 \gsim M^2$ the $pQCD$ moments $A_n(Q^2)$ are still leading, while resonances contribute to the higher-twist terms $B_{nk}(Q^2)$. Moreover, in Ref. \cite{RGP77} it was shown that the quantities $A_n(Q^2)$ are the Cornwall-Norton moments of a smooth structure function, which can be identified with the average function $F_2^{TM}(\xi, Q^2)$ occurring in the $BG$ local duality, namely:
 \be
      A_n(Q^2) = \int_0^1 d\xi ~ \xi^n {F_2^{TM}(\xi, Q^2) \over \xi^2}
     \label{eq:Mn_dual}
 \ee
with
 \be
      F_2^{TM}(\xi, Q^2) & = & {x^2 \over (1 + {4 M^2 x^2 \over Q^2})^{3/2}} 
      {F_2^S(\xi, Q^2) \over \xi^2} + 6 {M^2 \over Q^2} {x^3 \over (1 + {4 
      M^2 x^2 \over Q^2})^2} \int_{\xi}^1 d\xi' {F_2^S(\xi', Q^2) \over 
      \xi'^2} +
       \nonumber \\
      & + & 12 {M^4 \over Q^4} {x^4 \over (1 + {4 M^2 x^2
      \over Q^2})^{5/2}} \int_{\xi}^1 d\xi' \int_{\xi'}^1 d\xi" 
      {F_2^S(\xi", Q^2) \over \xi"^2}
      \label{eq:dual}
 \ee
where the $\xi$-dependence as well as the various integrals appearing in the r.h.s. account for target-mass effects in the $OPE$ of the hadronic tensor.  The latter appear as power corrections, which are however of {\em kinematical} origin, i.e. not related to partonic correlations. In Eq. (\ref{eq:dual}) $F_2^S(x, Q^2)$ represents the scaling structure function, fitted to high $Q^2$ data and extrapolated down to low values of $Q^2$ by the $pQCD$ evolution equations \cite{AP77}; namely, in terms of the parton distribution functions $q_f(x, Q^2)$ one has:
 \be
       F_2^S(x, Q^2) = \sum_f e_f^2 ~ x \left [ q_f(x, Q^2) + 
      \bar{q}_f(x, Q^2) \right ]
      \label{eq:F2_scaling}
 \ee
Note that the scaling function $F_2^S$ enters Eq. (\ref{eq:dual}) with its argument $x$ directly replaced by the Nachtmann variable $\xi$.

\indent Thus, the suggestion of Ref. \cite{RGP77}, applied to the analysis of available nucleon and nuclear inclusive data in Refs. \cite{duality,elastic}, is that the $BG$ duality manifests itself as the dominance of the leading-twist moments $A_n(Q^2)$ in the $Q^2$ behaviour of the low-order moments $M_n(Q^2)$, starting from a value $Q^2 \simeq Q_{BG}^2$ almost independent of the order $n$ of the moment\footnote{For high values of $n$ the moments of the structure functions $F_2(\xi, Q^2)$ and $F_2^{TM}(\xi, Q^2)$ should differ because of the rapidly varying behaviour of the nucleon-resonance peaks.}. At variance with the analyses of Refs. \cite{duality,elastic}, where the Cornwall-Norton and Nachtmann definitions of the moments have been considered, in this work truncated Cornwall-Norton moments, limited to the nucleon-resonance production regions, will be employed. This choice is worthwhile in view of the forthcoming high-precision data from $JLAB$ \cite{JLAB}, which are expected to cover the nucleon-resonance production regions.

\indent Our aim is twofold: ~ i) to clarify that the $BG$ duality does not follow from a compensation among dynamical and kinematical power corrections; in this respect, we will show that the main role played by target-mass corrections is to make the dual function (\ref{eq:dual}) only smoothly dependent on $Q^2$, so that the latter at low $Q^2$ can resemble the scaling curve seen at high $Q^2$; ~ ii) to show in a direct way that the concept of $BG$ duality cannot be extended to the nucleon elastic peak. As for the latter point, note that the inelastic part of the structure function $F_2(\xi, Q^2)$ is vanishing  beyond the pion inelastic threshold
 \be
       \xi_{\pi} \equiv {2 x_{\pi} \over 1 + \sqrt{1 + {4 M^2 x_{\pi}^2 
       \over Q^2}}}  \leq 1
      \label{eq:csi_pi}
\ee
where $x_{\pi} = Q^2 / [Q^2 + (M + m_{\pi})^2 - M^2]$, so that for $\xi_{\pi} \leq \xi \leq 1$ the function $F_2(\xi, Q^2)$ contains only the nucleon elastic peak contribution. Since the theoretical {\em dual} function (\ref{eq:dual}) is non-vanishing for $\xi_{\pi} \leq \xi \leq 1$, we will address the specific question\footnote{The author thanks Anatoly Radyushkin for drawing his attention to this point.} whether the contribution of the nucleon elastic peak to the moments (\ref{eq:Mn_CN}) may be dual to the unphysical part of the moments (\ref{eq:Mn_dual}) of the target-mass-corrected leading-twist structure function $F_2^{TM}(\xi, Q^2)$.

\section{The $Q^2$ behaviour of the dual function $F_2^{TM}(\xi, Q^2)$}

\indent In this work the scaling function $F_2^S(x, Q^2)$ is evaluated adopting the $GRV$ set of parton distributions functions \cite{GRV}, which can be  evolved at next-to-leading order ($NLO$) in the strong coupling constant down to $Q^2 \sim 0.4 ~ (GeV/c)^2$. Then, the target-mass corrected function $F_2^{TM}(\xi, Q^2)$ can be easily calculated via Eq. (\ref{eq:dual}) \footnote{Strictly speaking, Eq. (\ref{eq:dual}) can be used only for $Q > M$. Indeed, by inverting Eq. (\ref{eq:csi}) one gets $x = \xi / (1 - M^2 \xi^2 / Q^2)$. Therefore, for $Q > M$ the $\xi$-range extends up to $\xi_{max} = 1$, corresponding to unphysical, but finite values of $x > 1$. If $Q \leq M$, the maximum allowable value of $\xi$ is $\xi_{max} = Q / M$, corresponding to $x \to \infty$. Thus, for $Q \leq M$ the upper limit of the integrals appearing in Eq. (\ref{eq:dual}) is not $1$ but $Q / M$ and the function $F_2^{TM}(\xi, Q^2)$ is vanishing for $\xi \geq Q / M$ (see Fig. 1(a)).}.

\indent The $BG$ local duality of the structure function of proton, deuteron and light nuclei has been already illustrated in detail in Refs. \cite{duality,elastic}. Here, in Fig. 1, we have compared the theoretical {\em dual} function $F_2^{TM}(\xi, Q^2)$ with the proton pseudo-data, obtained adopting the $SLAC$ interpolation formula of Ref. \cite{SLAC} (used also in Refs. \cite{duality,elastic}), for few values of $Q^2$ corresponding to some of the kinematical conditions of the $JLAB$ experiment \cite{JLAB}. It can be seen that the onset of the $BG$ duality occurs at $Q^2 \simeq Q_{BG}^2 \sim 1.5 ~ (GeV/c)^2$ (see Refs. \cite{duality,elastic} for more details).

\indent Since the function $F_2^S$ enters Eq. (\ref{eq:dual}) with its argument $x$ replaced by $\xi$, the $Q^2$-dependence of the $\xi$-shape of $F_2^S(\xi, Q^2)$ is shown in Fig. 2(a) for values of $Q^2$ of interest in this work. It can be seen that at large $\xi$ the $pQCD$ evolution decreases systematically the scaling function $F_2^S(\xi, Q^2)$ as $Q^2$ increases. This is a well known feature related to the fact that high-momentum valence quarks are slowered by gluon bremsstrahlung.  Thus, the $\xi$-shape of the scaling function $F_2^S(\xi, Q^2)$ exhibits a non-negligible $Q^2$-dependence for $0.5 \lsim Q^2 ~ (GeV/c)^2 \lsim 10$.

\indent In Fig. 2(b) the results of the calculation of the target-mass corrected function $F_2^{TM}(\xi, Q^2)$ (see Eq. (\ref{eq:dual})) are plotted versus $\xi$ for various values of $Q^2$. It can clearly be seen that for $Q^2 \geq Q_{BG}^2 \sim 1.5 ~ (GeV/c)^2$ the shape of $F_2^{TM}(\xi, Q^2)$ at large $\xi$ is only slightly sensitive to $Q^2$, leading to the approximate relation
 \be
       F_2^{TM}(\xi, Q^2) \sim F_2^S(\xi, Q_{high}^2)
      \label{eq:BG_TM}
 \ee
valid for $\xi \gsim 0.2$ and $1.5 \lsim Q^2 ~ (GeV/c)^2 \lsim 10$ with $Q_{high}^2 \sim 30 ~ (GeV/c)^2$. The result (\ref{eq:BG_TM}) implies that the $\xi$-shape of the target-mass corrected function $F_2^{TM}(\xi, Q^2)$ at low $Q^2$, which thanks to local duality is representative of the $BG$ curve averaged over the resonance bumps (see Fig. 1), resembles quite closely the $\xi$-shape of the scaling function $F_2^S(\xi, Q^2)$ seen at high $Q^2$. In other words, in the window $1.5 \lsim Q^2 ~ (GeV/c)^2 \lsim 10$ the target-mass corrections almost compensate the effects due to the $pQCD$ evolution. 

\indent Thus, target-mass effects are extremely important for the occurrence of the $BG$ duality, though they are clearly not sufficient. Indeed, the inclusion of target-mass effects (and $pQCD$ evolution) as well as the smallness of the {\em dynamical} higher-twist effects, when locally averaged, are required for producing the observed $BG$ duality of the proton structure function.

\section{Truncated moments}

We want now to address the phenomenology of the $BG$ local duality adopting truncated Cornwall-Norton moments, limited to the nucleon-resonance production regions. To this end we introduce the following quantities
 \be
       \bar{M_n}(Q^2) \equiv \int_{\xi^*}^{\xi_{\pi}} d\xi ~ \xi^n {F_2(\xi, 
       Q^2) \over \xi^2} 
       \label{eq:Mn_res}
 \ee
and
 \be
       \bar{A_n}(Q^2) \equiv \int_{\xi^*}^{\xi_{\pi}} d\xi ~ \xi^n 
       {F_2^{TM}(\xi,  Q^2) \over \xi^2} 
      \label{eq:Mndual_res}
 \ee
where $\xi^* = 2 x^* / (1 + \sqrt{1 + 4 M^2 x^{*2} /Q^2})$ with  $x^* = Q^2 / [Q^2 + W^{*2} - M^2]$. In what follows we will adopt the value $W^* = 2.5 ~ GeV$ for a full coverage of the most prominent nucleon-resonance bumps. Note that for truncated moments an $OPE$ analogous to Eq. (\ref{eq:Mn_DGP}) cannot hold any more, and that truncated moments are expected to be more sensitive to the degree of locality of the $BG$ duality with respect to untruncated ones. Finally, we will consider also the truncated Cornwall-Norton moments of the scaling function $F_2^S(x, Q^2)$, viz.
 \be
       \bar{A_n^S}(Q^2) \equiv \int_{x^*}^{x_{\pi}} dx ~ x^n {F_2^S(x, Q^2) 
      \over x^2} 
       \label{eq:MnS_res}
 \ee
which do not contain target-mass corrections\footnote{When $M = 0$ (no target mass), one has $\xi \to x$, $F_2^{TM}(\xi, Q^2) \to F_2^S(x, Q^2)$ and therefore $\bar{A}_n(Q^2)$ (Eq. (\ref{eq:Mndual_res})) reduces to $\bar{A}_n^S(Q^2)$ (Eq. (\ref{eq:MnS_res})). When $M \neq 0$, the difference between $\bar{A}_n(Q^2)$ and $\bar{A}_n^S(Q^2)$ measures the relevance of the target-mass corrections in the truncated moments.}. 

\indent The moments (\ref{eq:Mn_res}) have been estimated using the $SLAC$ interpolation formula \cite{SLAC} and compared with the theoretical moments (\ref{eq:Mndual_res}) and (\ref{eq:MnS_res}) obtained starting from the $GRV$ set of parton distribution functions \cite{GRV}. We point out that: ~ i) the use of the interpolation formula of Ref. \cite{SLAC} allows to cover the ranges of $\xi$ and $Q^2$ which are crucial for the evaluation of the moments considered in this work; therefore, the uncertainties on the experimental moments are related only to the accuracy of the interpolation formula, which is simply given by a $\pm 4 \%$ total (systematic + statistical) error reported for the $SLAC$ data of Ref. \cite{SLAC}; ~ ii) in case of the proton the large-$x$ behaviour of the $GRV$ set has been found \cite{twist} to produce low-order untruncated moments which agree very nicely with the leading-twist terms extracted from a twist analysis of the proton world data.

\indent The results obtained for the ratios $\bar{M_n}(Q^2) / \bar{A_n}(Q^2)$ and $\bar{M_n}(Q^2) / \bar{A_n^S}(Q^2)$ are plotted in Fig. 3 for various values of $n \leq 10$. It can be seen that the truncated moments $\bar{A_n^S}(Q^2)$ largely deviates from  $\bar{M_n}(Q^2)$, while on the contrary the target-mass-corrected truncated moments $\bar{A_n}(Q^2)$ deviates only slightly from the pseudo-experimental moments for $Q^2 \geq Q_{BG}^2 \sim 1.5 ~ (GeV/c)^2$. This means that: ~ i) the effects of the kinematical power corrections are relevant in the $Q^2$-range of the present analysis\footnote{The use of a smaller upper limit $W^* = 2 ~ GeV$ in Eqs. (\ref{eq:Mn_res}) and (\ref{eq:MnS_res}) makes the ratio $\bar{M_n}(Q^2) / \bar{A_n^S}(Q^2)$ even larger than the one shown in Fig. 3.}, and ~ ii) for $Q^2 \geq Q_{BG}^2 \sim 1.5 ~ (GeV/c)^2$ the effects of {\em dynamical} higher twists reduce only to a $\sim 20 \div 30 \%$ deviation from the (target-mass corrected) leading-twist contribution. Note that the onset of the $BG$ duality, indicated by a sharp decrease of the slope of $\bar{M_n}(Q^2) / \bar{A_n}(Q^2)$ (see Fig. 3), occurs at a value $Q^2 \simeq Q_{BG}^2$ which is almost independent of the order $n$ of the moment.

\indent To sum up, the analysis of the $Q^2$ behaviour of low-order truncated moments, limited to the nucleon-resonance production regions, shows the dominance of the leading-twist term starting from $Q^2 \simeq Q_{BG}^2 \sim 1.5 ~ (GeV/c)^2$, in analogy with the results obtained using untruncated moments in Refs. \cite{duality,elastic}. We stress however that the $BG$ local duality seen in terms of truncated moments cannot have any $OPE$-based justification.

\section{Elastic peak contribution}

The contribution of the nucleon elastic peak to the moments (\ref{eq:Mn_CN}) reads as (cf. Ref. \cite{duality})
 \be
       M_n^{el}(Q^2) & \equiv & \int_{\xi_{\pi}}^1 d\xi ~ \xi^n 
      {F_2^{el}(\xi, Q^2) \over \xi^2} = {[G_E(Q^2)]^2 + \tau [G_M(Q^2)]^2 
      \over 1 + \tau} \int_{\xi_{\pi}}^1 d\xi ~ \xi^{n-2} \delta(x - 1)
      \nonumber \\
      & = & {[G_E(Q^2)]^2 + \tau [G_M(Q^2)]^2 \over 1 + \tau} {\xi_{el}^n 
      \over 2 - \xi_{el}}
      \label{eq:Mn_el}
 \ee
where $\tau \equiv Q^2 / 4M^2$ and $\xi_{el} \equiv 2 / (1 + \sqrt{1 + 1 / \tau})$. In the spirit of the $BG$ local duality and of the analysis performed previously in  terms of truncated moments, it seems reasonable to ask whether the term (\ref{eq:Mn_el}) may be dual to the corresponding integral of the structure function (\ref{eq:dual}) in the unphysical region $\xi_{\pi} \leq \xi \leq 1$, i.e. if $M_n^{el}(Q^2)$ may be dual to
 \be
      A_n^{unphys}(Q^2) \equiv \int_{\xi_{\pi}}^1 d\xi ~ \xi^n 
     {F_2^{TM}(\xi, Q^2) \over \xi^2}
     \label{eq:Mndual_el}
 \ee
Such a super-local version of the $BG$ duality would be highly non-trivial, since only one hadronic channel (i.e., the nucleon elastic peak) would be involved. Moreover, if the local duality would start already at $Q^2 \simeq Q_{BG}^2 \sim 1.5 ~ (GeV/c)^2$, one would be able to extract the nucleon form factor at low and intermediate values of $Q^2$ from the large-$x$ behaviour of the parton distribution functions. More precisely, if $M_n^{el}(Q^2) \simeq A_n^{unphys}(Q^2)$, the magnetic form factor $G_M(Q^2)$ could be extracted independently of $n$ via the following equation
 \be
      [G_M(Q^2)]^2 \simeq  {2 - \xi_{el} \over \xi_{el}^n} A_n^{unphys}(Q^2) 
      ~ \mu^2 {1 + \tau \over 1 + \mu^2 \tau}
      \label{eq:GM}
 \ee
where we have further assumed the scaling-law behaviour $G_E(Q^2) \simeq G_M(Q^2) / \mu$, with $\mu \equiv G_M(0)$ being the nucleon magnetic moment.

\indent We have evaluated Eq. (\ref{eq:Mndual_el}) starting from the $GRV$ set of parton distribution functions \cite{GRV} and we have extracted the form factor $G_M(Q^2)$ in case of the proton via Eq. (\ref{eq:GM}). The values obtained for $G_M^p(Q^2)$ are plotted in Fig. 4 and compared with the experimental data of Ref. \cite{data}. It can clearly be seen that: ~ i)  the r.h.s. of Eq. (\ref{eq:GM}) is almost totally independent of $n$, at least for $2 \leq n \leq 10$ and $Q^2 \geq Q_{BG}^2 \sim 1.5 ~ (GeV/c)^2$, and ~ ii) the values of $G_M^p(Q^2)$ extracted assuming local duality are well below the experimental points by a factor of $2 \div 3$.

\indent Thus, the occurrence of a super-local version of the $BG$ duality in the elastic peak region is found to be very critical at low and intermediate values of $Q^2$. The same conclusion was suggested also by the analysis of Ref. \cite{elastic}, carried out in terms of untruncated Nachtmann moments covering the full kinematical $x$-range.

\indent The observation that the $BG$ duality holds in the nucleon-resonance regions, but not for the nucleon elastic peak, strongly suggests to interpret the $BG$ duality as a non-trivial compensation among local averages of dynamical higher-twist effects in the nucleon-resonance bumps and in the background under them. Note that a close relation between the $Q^2$-behaviour of the background and the $N - \Delta(13232)$ transition peak has been already pointed out in Ref. \cite{CM93}.

\section{Conclusions}

\indent The occurrence of the Bloom-Gilman local duality in the low-order moments of the nucleon structure function has been investigated for values of the squared four-momentum transfer $Q^2$ between $\sim 0.5$ and $\sim 10 ~ (GeV/c)^2$.  At variance with previous analyses truncated Cornwall-Norton moments, limited to the nucleon-resonance production regions, have been considered.

\indent The role played by target-mass corrections has been illustrated, showing that target-mass effects are necessary (but not sufficient) for producing the observed Bloom-Gilman duality of the nucleon structure function. In particular, it has been shown that in the window $1.5 \lsim Q^2 ~ (GeV/c)^2 \lsim 10$ target-mass corrections almost compensate the effects due to the $pQCD$ evolution. Moreover, the $Q^2$ behaviour of the truncated moments exhibits the phenomenology of the Bloom-Gilman local duality in good agreement with the results obtained adopting untruncated moments \cite{duality,elastic}. 

\indent The possibility of a super-local version of the Bloom-Gilman duality between the {\em unphysical} region at large values of the Nachtmann variable and the nucleon elastic peak contribution has been analyzed. It has been found that the proton magnetic form factor extracted assuming local duality is significantly below the experimental data at low and intermediate values of $Q^2$.

\section*{Acknowledgments}

The author gratefully acknowledges R. Ent, C. Keppel and A. Radyushkin for fruitful comments and discussions.

\newpage

\begin{figure}[htb]

\centerline{\epsfxsize=14cm \epsfig{file=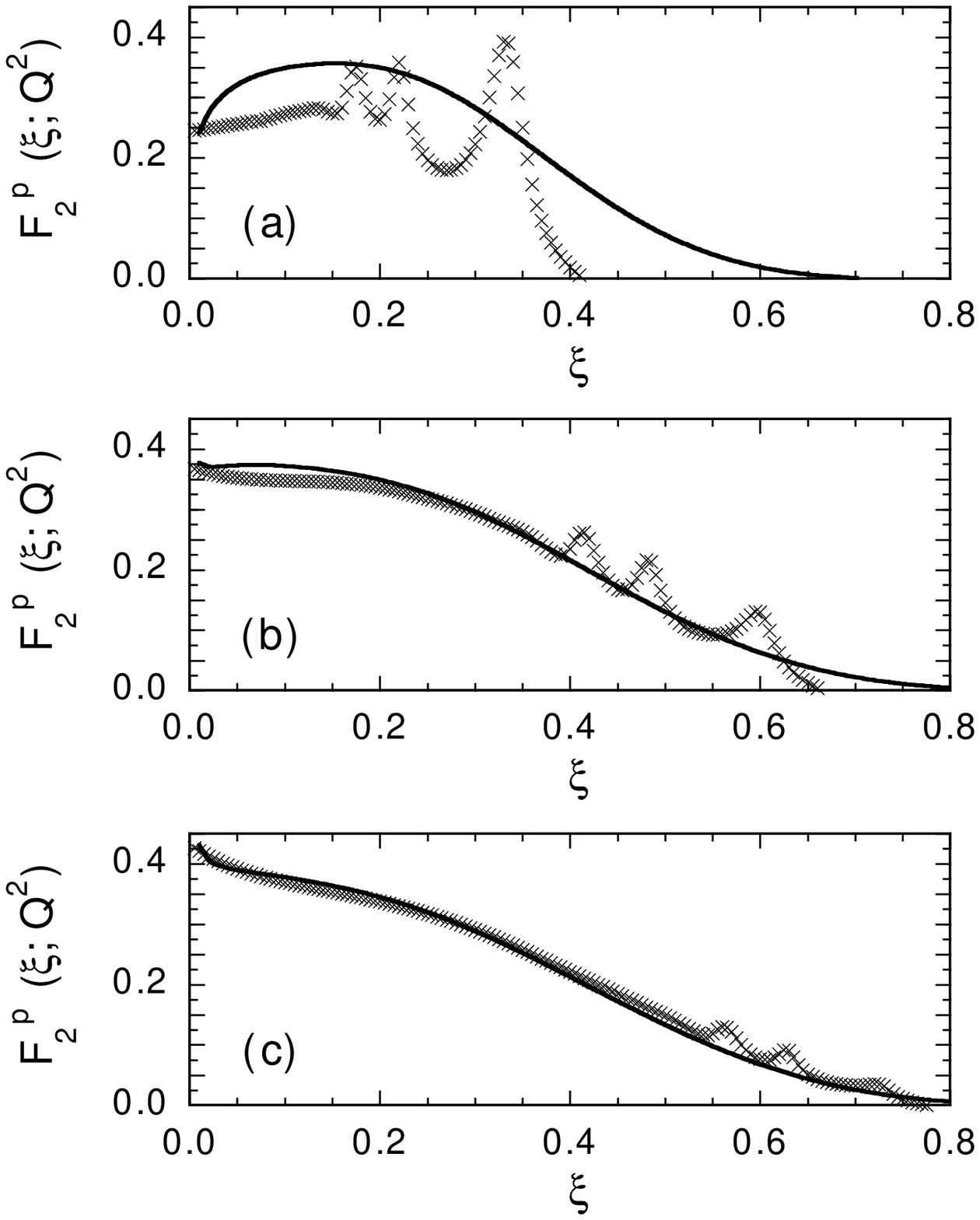}}

\vspace{0.5cm}

\small{{\bf Figure 1.} The proton structure function $F_2^p(Q^2)$ (crosses), 
estimated via the $SLAC$ interpolation formula of Ref. \cite{SLAC}, versus 
the Nachtmann variable $\xi$ (see Eq. (\ref{eq:csi})). In (a), (b) and (c) 
the value of $Q^2$ is  $0.45, ~ 1.7$ and $3.3 ~ (GeV/c)^2$, respectively. 
The solid lines are the results of the calculations of the dual function 
(\ref{eq:dual}), obtained starting from the $GRV$ set of parton distribution 
functions of Ref. \cite{GRV}.}

\end{figure}

\newpage

\begin{figure}[htb]

\centerline{\epsfxsize=18cm \epsfig{file=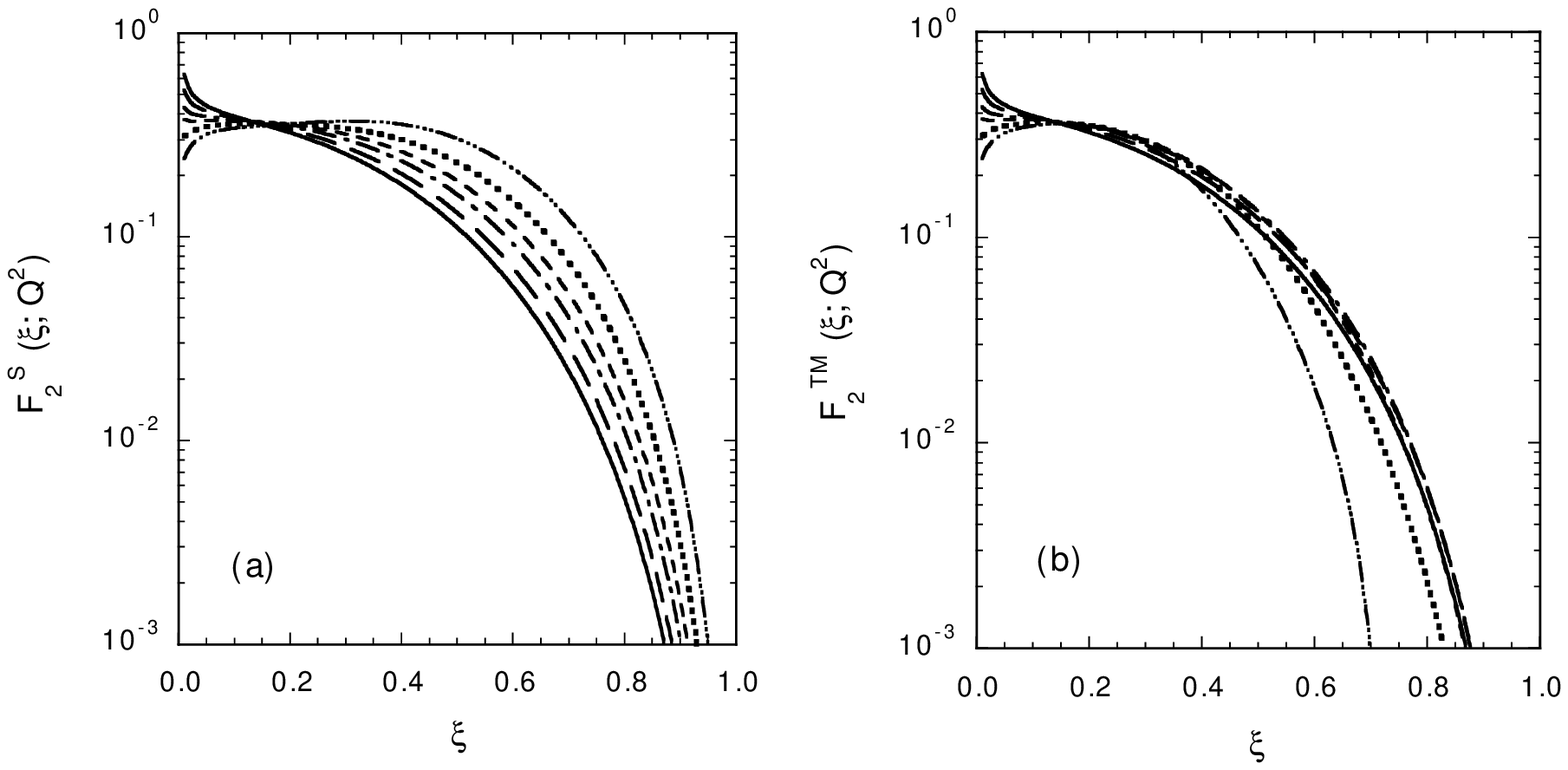}}

\vspace{0.5cm}

\small{{\bf Figure 2.} The (proton) scaling function $F_2^S(\xi, Q^2)$ (a) 
and the (proton) target-mass-corrected function $F_2^{TM}(\xi, Q^2)$ (b) 
versus $\xi$ for various values of $Q^2$. The triple-dotted-dashed, dotted, 
short-dashed, dot-dashed, long-dashed and solid lines correspond to $Q^2 = 
0.45, 0.85, 1.7, 3.3, 10$ and $30 ~ (GeV/c)^2$, respectively. The $GRV$ set 
of parton distribution functions \cite{GRV} has been adopted.}

\end{figure}

\newpage

\begin{figure}[htb]

\centerline{\epsfxsize=16cm \epsfig{file=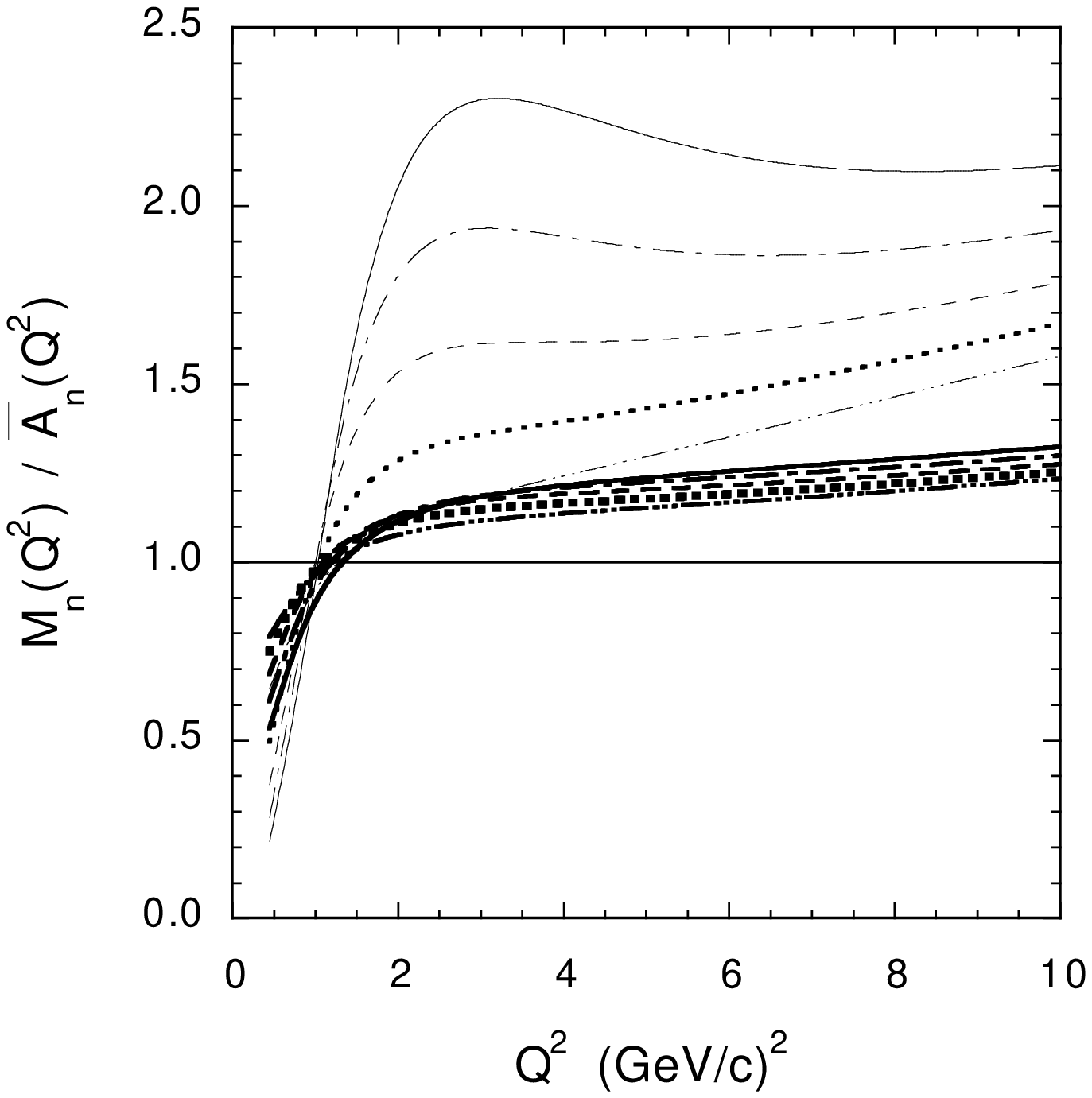}}

\vspace{0.5cm}

\small{{\bf Figure 3.} The ratios $\bar{M_n}(Q^2) / \bar{A_n}(Q^2)$ (thick 
lines) and $\bar{M_n}(Q^2) / \bar{A_n^S}(Q^2)$ (thin lines) obtained using 
Eqs. (\ref{eq:Mn_res}-\ref{eq:MnS_res}) versus $Q^2$ (see text). The 
triple-dotted-dashed, dotted, short-dashed, dot-dashed and solid lines 
correspond to $n = 2, 4, 6, 8$ and $10$, respectively.}

\end{figure}

\newpage

\begin{figure}[htb]

\centerline{\epsfxsize=16cm \epsfig{file=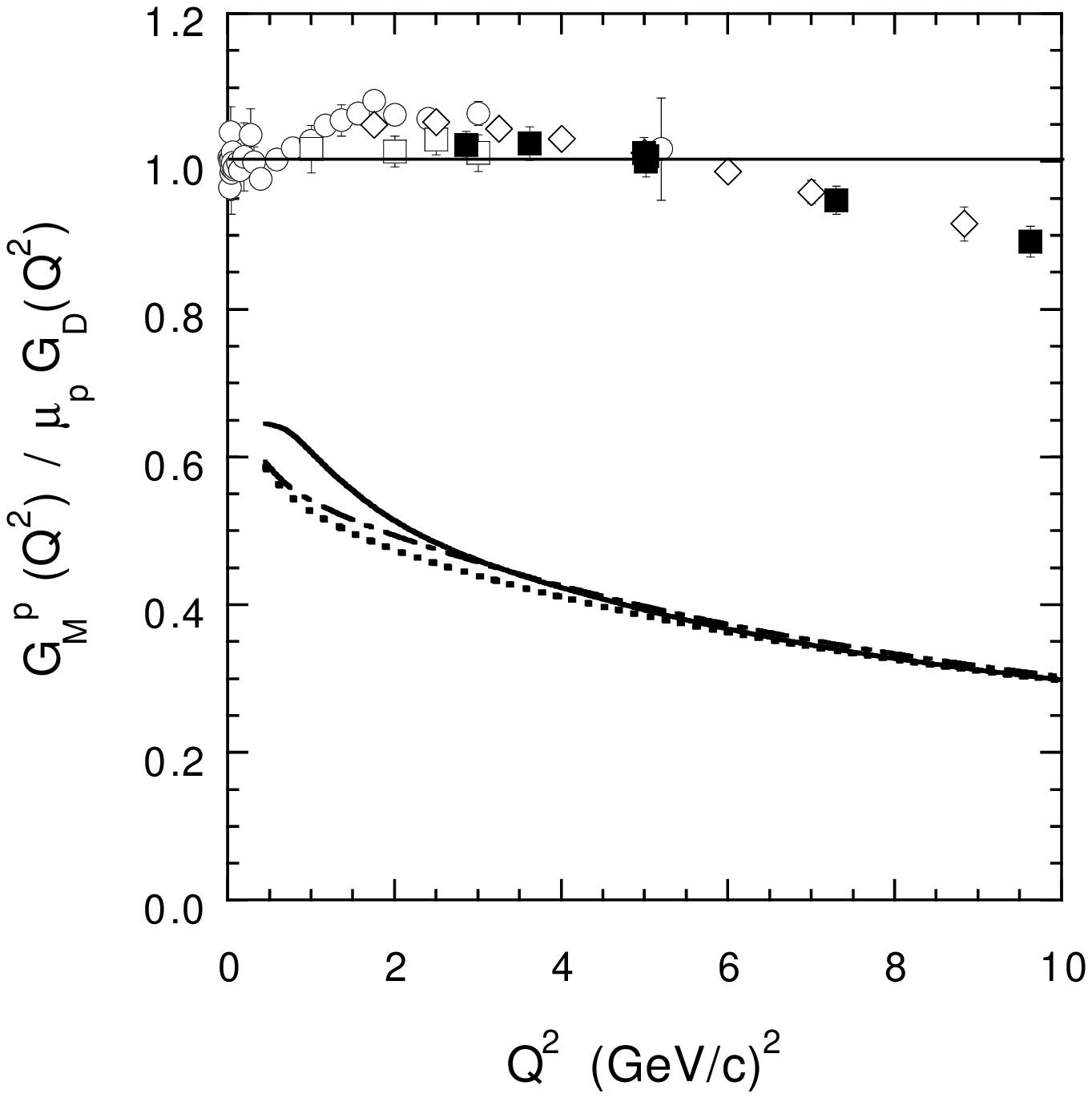}}

\vspace{0.5cm}

\small{{\bf Figure 4.} The ratio of the proton magnetic form factor 
$G_M^p(Q^2)$ to the dipole ans\"atz $\mu_p G_D(Q^2) = 2.793 / (1 + Q^2 / 
0.71)^2$ versus $Q^2$. Open dots, squares, diamonds and full squares are 
the experimental data from Ref. \cite{data}(a), (b), (c) and (d), 
respectively. The dot-dashed and solid lines are the results of Eq. 
(\ref{eq:GM}) at $n = 2$ and $n = 10$, respectively. The dotted line 
corresponds to the use of Eq. (\ref{eq:GM}) with $n = 2$, but including 
the effects of the suppression of the ratio $G_E^p(Q^2) / G_M^p(Q^2)$ 
with respect to the scaling-law expectation $G_E^p(Q^2) / G_M^p(Q^2) = 
1 / \mu_p$, observed recently at $JLAB$ \cite{ratio}.}

\end{figure}

\end{document}